\pdfoutput=1
\documentclass[conference]{IEEEtran}
\IEEEoverridecommandlockouts
\usepackage{cite}
\usepackage{bbm}
\usepackage{amsmath,amssymb,amsfonts}
\usepackage{algorithmic}
\usepackage{graphicx}
\usepackage{textcomp}
\usepackage{xcolor}
\usepackage[english]{babel}
\usepackage[acronym,nonumberlist]{glossaries}
\usepackage{glossaries-prefix}
\usepackage{pgfplots}
\usepgfplotslibrary{fillbetween}
\usepackage{amssymb}
\usepackage{import}
\usepackage{bm}
\usepackage{siunitx}
\usepackage{subcaption}

\usepackage[inline]{enumitem}

\usepackage[capitalise]{cleveref}

\newacronym[shortplural=GMMs]{GMM}{GMM}{Gaussian mixture model}
\newacronym[shortplural=HMMs]{HMM}{HMM}{hidden Markov model}
\newacronym[shortplural=DNNs]{DNN}{DNN}{deep neural network}
\newacronym[shortplural=SVDs]{SVD}{SVD}{singular value decomposition}
\newacronym[
    prefixfirst={a\ },
    prefix={an\ }
]{MCTS}{MCTS}{Monte Carlo tree search}
\newacronym[prefixfirst={a\ },prefix={an\ }]{MDP}{MDP}{Markov decision process}
\newacronym{CMDP}{CMDP}{constrained Markov decision process}
\newacronym{RL}{RL}{reinforcement learning}
\newacronym[shortplural=DTs]{DT}{DT}{decision tree}
\newacronym{SMT}{SMT}{satisfiability modulo theories}
\newacronym{IL}{IL}{Imitation Learning}
\newacronym[shortplural=CNNs]{CNN}{CNN}{convolutional neural network}
\newacronym[shortplural=DQNs]{DQN}{DQN}{deep Q-network}
\newacronym{AI}{AI}{artificial intelligence}
\newacronym{PPO}{PPO}{proximal policy optimization}
\newacronym{ML}{ML}{machine learning}
\newacronym{QML}{QML}{quantum machine learning}
\newacronym{NISQ}{NISQ}{noisy intermediate scale quantum}
\newacronym{QC}{QC}{quantum circuit}
\newacronym{VQC}{VQC}{variational quantum circuit}
\newacronym{MNIST}{MNIST}{modified national institute of standards and technology}
\newacronym{FIM}{FIM}{Fisher information matrix}
\newacronym{IDU}{IDU}{incremental data-uploading}
\newacronym{DRU}{DRU}{data re-uploading}
\newacronym{QRL}{QRL}{quantum reinforcement learning}

\usepackage{lineno,todonotes}

\begin{document}

\title{Incremental Data-Uploading \\ for Full-Quantum Classification
\thanks{
The research is supported by the Bavarian Ministry of Economic Affairs, Regional Development and Energy with funds from the Hightech Agenda Bayern.\\
email address for correspondence: \\
maniraman.periyasamy@iis.fraunhofer.de, axel.plinge@iis.fraunhofer.de}
}

\author{
\IEEEauthorblockN{Maniraman Periyasamy, Nico Meyer, Christian Ufrecht,
Daniel D.\ Scherer, Axel Plinge, and Christopher Mutschler}
\IEEEauthorblockA{\textit{Fraunhofer IIS, Fraunhofer Institute for Integrated Circuits IIS},
Nuremberg, Germany \\\vspace{1mm}}
}

\maketitle

\begin{abstract} 
The data representation in a machine-learning model strongly influences its performance. This becomes even more important for quantum machine learning models implemented on noisy intermediate scale quantum (NISQ) devices. Encoding high dimensional data into a quantum circuit for a NISQ device without any loss of information is not trivial and brings a lot of challenges. While simple encoding schemes (like single qubit rotational gates to encode high dimensional data) often lead to information loss within the circuit, complex encoding schemes with entanglement and data re-uploading lead to an increase in the encoding gate count. This is not well-suited for NISQ devices. This work proposes `incremental data-uploading', a novel encoding pattern for high dimensional data that tackles these challenges. We spread the encoding gates for the feature vector of a given data point throughout the quantum circuit with parameterized gates in between them. This encoding pattern results in a better representation of data in the quantum circuit with a minimal pre-processing requirement. We show the efficiency of our encoding pattern on a classification task using the MNIST and Fashion-MNIST datasets, and compare different encoding methods via classification accuracy and the effective dimension of the model.
\end{abstract}


\glsresetall
\section{Introduction}
\label{sec:introduction}

\IEEEPARstart{T}{he} field of quantum computing has witnessed a surge of interest from different fields in the scientific community recently. With the realization of quantum devices with increasing qubits and fidelity by several manufacturers like IBM~\cite{ibmq2022} and Google~\cite{Arute2019}, research in quantum computing started shifting from theoretical research towards applied research~\cite{Bova2021}. Quantum devices employed for information processing and computational purposes are currently being explored to overcome many of the limitations posed by classical hardware in different industry segments such as finance, cybersecurity, and chemical industry~\cite{Bova2021}. Research groups from diverse fields of science and technology are studying numerous heuristic and non-heuristic quantum computing methods to solve a given problem and achieve the so-called quantum supremacy~\cite{Moussa2020}.

However, the current limitation in the number of qubits and low gate fidelity makes non-heuristic approaches impractical. Hence heuristic approaches, especially in the domain of \gls{ML}, are deemed to be one of the prime candidates for practical quantum computing in the NISQ era. 
Though classical \gls{ML} is well matured and has decades of domain-specific enhancements, its quantum counterpart is still a lively research field with many loose ends and uncertainties~\cite{Nimish2021QMLreview}. Adding in the factors from NISQ devices like low gate fidelity and qubit connectivity limitations to this mixture of uncertainties, learning-based approaches like \gls{QML} and \gls{QRL} become more complex.

\begin{figure}[t!]
\centering
\tiny
\def\svgwidth{\columnwidth}
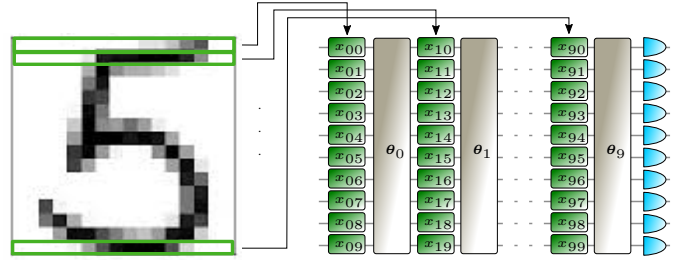
\caption{Proposed method: The image is downscaled to $10\times 10$, then fed row by row to the 10 bit quantum circuit, each encoding layer (green) is followed by a variational layer (gray), then the 10 qubits are measured (blue).}
\label{fig:overview}
\end{figure}

Of all the open questions yet to be answered in the field of gate based quantum computing and quantum machine learning, the question of selecting the optimal \gls{VQC} for a given problem is of utmost importance and significance~\cite{Watanabe2021}. The gates in a \gls{QC} used for \gls{QML} are grouped into three categories, namely, encoding gates, decoding gates and the variational gates. These encoding gates, decoding gates and variational gates are further grouped into encoding layer, decoding layer and variational layer in many \gls{QML} works, though these layers are just a visual representation and do not reflect the theory behind layers in classical ML. The encoding gates are selected based on the encoding method chosen and the number of input features. The optimal selection of the encoding method is pivotal to successful learning of a \gls{QML} model as this represents the classical information to be fed into the circuit~\cite{yano2020efficient, Caro2021encodingdependent, LaRose2020, Banchi2021}. While the effect of different encoding methods in data representation and expressivity of a \gls{VQC} have been studied before~\cite{Abbas_2021, Schuld_2021, Caro2021, Banchi2021, franz2022uncovering}, the encoding pattern we propose (see \cref{sec:method}) has not been discussed in the literature previously to the best of our knowledge.

The encoding gate set positioning becomes a salient factor for \gls{VQC}-based \gls{QML} models that learn from high dimensional data due to the limitations in the size of the quantum device in terms of the number of qubits and the depth of the QC that can be executed both in a simulation and on a real device. This limitation brings in a trade-off between the number of encoding gates and the number of parameterized gates for a fixed circuit depth and gate count. The larger the dimension of the input, the larger the number of encoding gates required and the larger the number of encoding gates used, the fewer the number of parameterized gates that can be used. The reduction in the parameterized gate count reduces the expressivity of the model. This reduction of the dimension of the input results in information loss. Naively encoding the data at the start of the circuit or encoding patterns like data re-uploading is not very promising for high dimensional data as it results in an increase in the QC depth and gate count or the information in data becomes less accessible by the model.

This paper investigates the impact of encoding gate set positioning on the trainability of a \gls{QML} model and proposes an encoding pattern for high dimensional inputs, see~\cref{fig:overview}. The key concept behind our method is incremental uploading.

We evaluate our approach on an image classification task (i.e., MNIST~\cite{MNIST} and Fashion MNIST~\cite{FashionMNIST}) as image inputs are among the most common high dimensional inputs with various practical significance. While classifying MNIST with \gls{QML} is not new, i.e., early work uses heavily downscaled $4 \times 4$ images for binary classification~\cite{farhi2018classification} or quantum techniques for dimensionality reduction and classification~\cite{Kerenidis2020SlowFeat}, a truncation of the dataset or extreme reduction in dimensions using a dimensionality reduction will only work for simple classification tasks which tolerate these information losses. One other approach which handles high dimensional image data effectively is the quantum convolutional neural network~\cite{Matic2022}. However, this is not a pure quantum approach, and the input image is broken into smaller pieces and fed into the circuit sequentially, resulting in a longer runtime, however, with the prospect for parallelization To show how to handle more complex classification tasks, we use a high dimensional representation of the full MNIST dataset.
\section{Method}
\label{sec:method}

Quantum gates in a \gls{VQC} can be grouped into three categories: encoding layers, decoding layers and variational layers. Due to the limitation in circuit depth and to avoid possible barren plateau effects~\cite{mcclean2018}, the \gls{QC} has to be designed in such a way that it allows for maximum classification performance and expressivity for a given gate count and circuit depth. To this end, we studied the effect of encoding layer positioning on the performance of a \gls{QML} model by decomposing them into smaller encoding layers and progressively increasing the number of variational parameters between the layers in the \gls{VQC}. Also, we would like to introduce the nomenclature used for grouping the encoding gates throughout this paper. From here on, the collection of all encoding gates is to be called an encoding block and the encoding block split in to a smaller group of encoding gates are to be called as encoding layer.

\begin{figure*}
{\tiny \centering
\def\svgwidth{\textwidth}
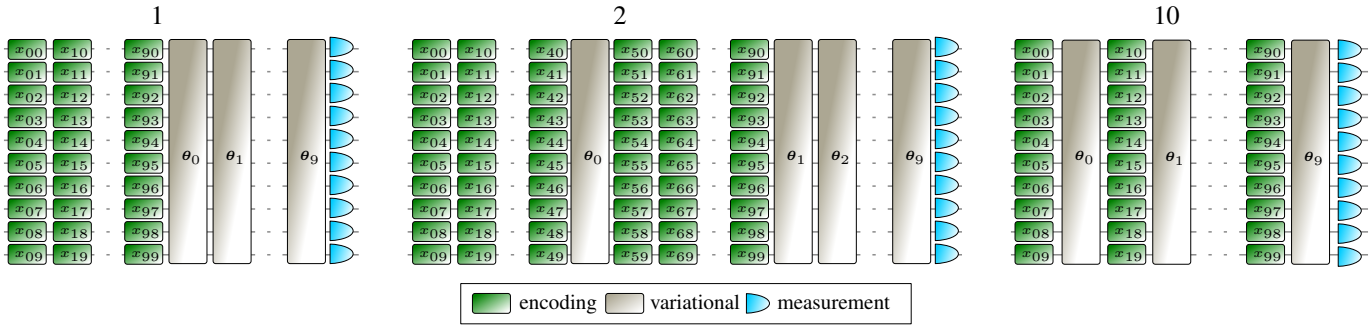}
%
\caption{Different splits of interleaved layers. For the 1-split, there are 10 encoding layers followed by 10 variational layers, for the 2-split, there are 5 encoding layers, followed by 1 variational layer, then 5 encoding layers, followed by the remaining 9 variational layers. The proposed 10-split is interleaving encoding and variational layers.}
\label{fig:splits}
\end{figure*}

An obvious decomposition of encoding block for image data splits and groups the gates used to encode raw features from each row of the input image. These row-wise grouped encoding gates (hereafter referred to as \textit{encoding layers}) can per design choice be freely moved across the \gls{VQC}, though each move results in a different architecture with an impact on the performance of the model. Hence, we designed five different encoding block split patterns with incremental number parameterized gates between them. These circuits are as follows: IDU\_1, here, there is no variational layer between the encoding layers. All encoding gates are placed at the start followed by all variational layers.  IDU\_2, IDU\_4, IDU\_8, IDU\_10 represent the circuits where the encoding block is split into 2, 4, 8, and 10 parts respectively. Between each split, there is a variational layer and the remaining variational layers are appended  at the end. The number of variational layers between any two encoding layers is restricted to one as we wanted to analyze the performance boost attained by introducing a minimal and constant number of variational layers in between them. This restriction is only a design choice and other design choices are of course possible. The overall working of this proposed encoding pattern is shown in~\cref{fig:splits}. We call this encoding pattern \gls{IDU}. The performance of this pattern is compared against the \gls{DRU} encoding pattern~\cite{Salinas2020} which is deemed to be the state-of-the-art following the evaluation metrics of Skolik et al.~\cite{Skolik2021} and theoretical support from Schuld et al.~\cite{Schuld_2021}. However, to have a fair comparison, the number of parameters of the \gls{QML} model is kept constant. Hence in the \gls{DRU} architecture, the entire image information is encoded into the circuit followed by a variational layer and this is repeated until the number of variational parameters matches the number of parameters used in the incremental data-uploading experiment.

\section{Evaluation}

\subsection{Data Statistics}
\label{sec:dataset}

Our experiments have been conducted using the MNIST~\cite{MNIST} and Fashion-MNIST~\cite{FashionMNIST} datasets. MNIST is a handwritten digit dataset consisting of 70,000 images representing the digits 0-9 with a size of $28\times 28$ pixels each. Each digit class contains roughly between 5,400 - 6,750 images. The dataset of 70,000 grayscale images has been randomly grouped into 48,000 images for training, 12,000 images for validation, and 10,000 images for evaluation. As our quantum experiments use Tensorflow Quantum and the Cirq simulator on classical hardware, deeper and larger circuits become computationally intractable. To reduce computation time we reduce the size of the images from $28\times 28$ to $10\times 10$ using a bilinear filter so that the encoding gate count required to encode the data is small. Similarly, Fashion MNIST is also made of 70,000 grayscale images of size $28\times 28$ representing 10 categories of clothes. As for MNIST we reduce the image size and hence the overall computational time.

\subsection{Quantum Encoding, Variational and Decoding Layer}

All our datasets consist of 10 classes. Hence, we designed a ten qubit quantum circuit along with the softmax function to learn a mapping function $f(\cdot): X \rightarrow R^{o}$, where $X$ is the dataset with $n$ data points and $R^{o}$ is the probability of a data point belonging to each class. The quantum circuit consists of multiple encoding and variational layers as explained below.

The process of embedding a classical data point $x \in X$ into a quantum circuit is commonly known as data encoding, sometimes also referred to as data uploading~\cite{Salinas2020}. In practice, one of the most common ways to encode a data point into a quantum circuit is via a state preparation circuit acting on state $|0\rangle^{\otimes n}$ in computational basis~\cite{LaRose2020}. A state preparation circuit often consists of single qubit rotational gates matching the dimension of $x$ with or without entangling gates so that each raw feature of the $x$ can be scaled between $\left[0, \pi\right]$ or $\left[0, 2\pi\right]$ and used as the rotational angles for one gate in the state preparation circuit~\cite{LaRose2020}. We choose a total of 100 single qubit rotational gates $R_x$ that match the feature dimension of each data point $x$ as the encoding layer(s) for all our experiments. The $R_x$ gates are split into groups of ten where each group acts on one qubit. Each pixel value in $x$, ranging between $\left[0, 255\right]$ is scaled to $\left[0, \pi\right]$ and are fed as the rotational angle for the encoding gates.

\begin{figure}[hbtp!]
    \centering
    \includegraphics[width=.95\linewidth]{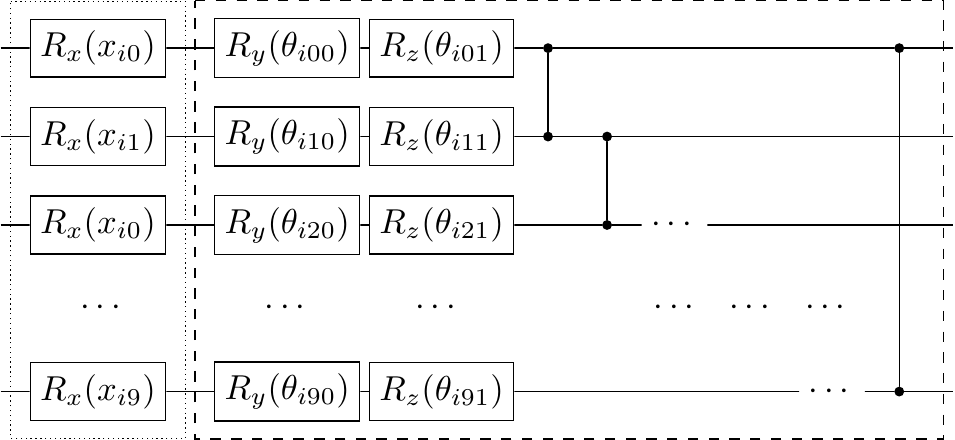}
    \caption{Single encoding block followed by a variational layer. The encoding block and the variational block are repeated 10 times with different intervals in between to form different architectures.}
    \label{fig:varialtional_layer}
\end{figure}

The variational layers hold the learnable parameters that are optimized using gradient descent to approximate the mapping function $f$. In quantum circuits, the variational layers are again realized using single-qubit rotational and multi-qubit entangling gates where the rotational angles of the rotational gates act as the learnable parameters. Our variational layers consist of single-qubit $R_y$ and $R_z$ rotational gates with nearest neighbour controlled-$R_z$ entanglements. The complete quantum circuit with encoding and variational layers is shown in~\cref{fig:varialtional_layer}. The circuit is measured in the computational basis, and the expectation values of the individual qubit along with the softmax function are used for class prediction.

\subsection{Incremental Data Uploading}

We implement our interval uploading method with each of the architectures representing a different split in the encoding layer. We train each architecture on all datasets with a learning rate of 0.001 using the ADAM optimizer~\cite{ADAM} for 25 epochs.

\begin{figure}
\begin{tikzpicture}
\begin{axis}[name=plot1,title={MNIST digits},
width=\linewidth, height=.3\textheight,
ylabel={validation accuracy [\%]},
xlabel={epoch},
xmin=1,xmax=20.5,
ymin=25,
ymax=68.5,
grid=major,
minor y tick  num=4,minor x tick  num=4,
tick label style={font=\footnotesize},
axis x line=bottom, axis y line=left, tick align = outside,
legend columns=-1,
legend style={/tikz/every even column/.append style={column sep=0.1cm},at={(0.5,1)},anchor=south,yshift=-5mm}, %
]

\addplot[draw=red, mark=o] table[draw=red, x=epoch,y=vamean] {results/compiled/arch_mnist_DRU.dat};
\addlegendentry{DRU}
\addplot[draw=blue, mark=+] table[x=epoch,y=vamean] {results/compiled/arch_mnist_1.dat};
\addlegendentry{1}
\addplot[draw=green,mark=triangle] table[ x=epoch,y=vamean] {results/compiled/arch_mnist_2.dat};
\addlegendentry{2}
\addplot[draw=yellow,mark=star] table[ x=epoch,y=vamean] {results/compiled/arch_mnist_4.dat};
\addlegendentry{4}
\addplot[draw=pink,mark=pentagon] table[ x=epoch,y=vamean] {results/compiled/arch_mnist_8.dat};
\addlegendentry{8}
\addplot[draw=black,mark=*] table[ x=epoch,y=vamean] {results/compiled/arch_mnist_10.dat};
\addlegendentry{10}

\addplot [name path=upper1,draw=none] table[x=epoch,y expr=\thisrow{vamean}+\thisrow{vastd}] {results/compiled/arch_mnist_1.dat};
\addplot [name path=lower1,draw=none] table[x=epoch,y expr=\thisrow{vamean}-\thisrow{vastd}] {results/compiled/arch_mnist_1.dat};
\addplot [draw=blue, fill=blue!10] fill between[of=upper1 and lower1];

\addplot [name path=upper2,draw=none] table[x=epoch,y expr=\thisrow{vamean}+\thisrow{vastd}] {results/compiled/arch_mnist_2.dat};
\addplot [name path=lower2,draw=none] table[x=epoch,y expr=\thisrow{vamean}-\thisrow{vastd}] {results/compiled/arch_mnist_2.dat};
\addplot [draw=green, fill=green!10] fill between[of=upper2 and lower2];

\addplot [name path=upper4,draw=none] table[x=epoch,y expr=\thisrow{vamean}+\thisrow{vastd}] {results/compiled/arch_mnist_4.dat};
\addplot [name path=lower4,draw=none] table[x=epoch,y expr=\thisrow{vamean}-\thisrow{vastd}] {results/compiled/arch_mnist_4.dat};
\addplot [draw=yellow, fill=yellow!10] fill between[of=upper4 and lower4];

\addplot [name path=upper8,draw=none] table[x=epoch,y expr=\thisrow{vamean}+\thisrow{vastd}] {results/compiled/arch_mnist_8.dat};
\addplot [name path=lower8,draw=none] table[x=epoch,y expr=\thisrow{vamean}-\thisrow{vastd}] {results/compiled/arch_mnist_8.dat};
\addplot [draw=pink, fill=pink!10] fill between[of=upper8 and lower8];

\addplot [name path=upper10,draw=none] table[x=epoch,y expr=\thisrow{vamean}+\thisrow{vastd}] {results/compiled/arch_mnist_10.dat};
\addplot [name path=lower10,draw=none] table[x=epoch,y expr=\thisrow{vamean}-\thisrow{vastd}] {results/compiled/arch_mnist_10.dat};
\addplot [draw=black, fill=black!10] fill between[of=upper10 and lower10];

\end{axis}

\end{tikzpicture}
\caption{Average validation accuracy and its standard deviation over 5 training runs for quantum circuits with different splits in the encoding layer. DRU stands for the Data Re-uploading method, and the numbers 1, 2, 4, 8, 10 represents the quantum circuits with 1 whole encoding layer, encoding layer split into 2, 4, 8, 10 blocks respectively.}
\label{fig:val_enc}
\end{figure}
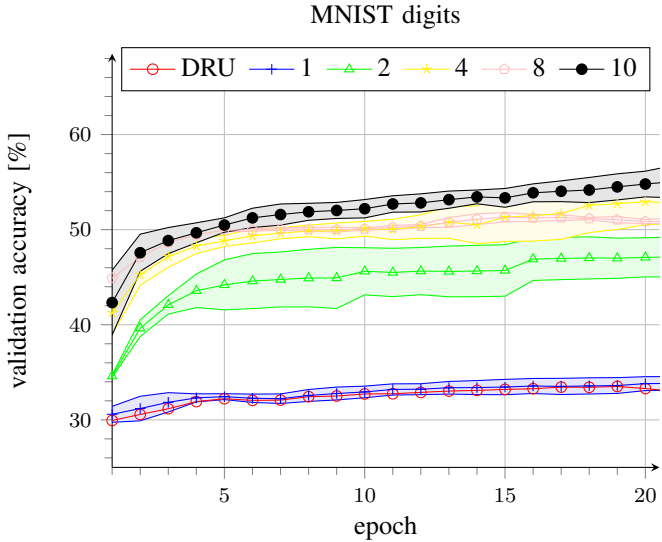

From \cref{fig:val_enc}, we can infer that there is a direct correlation between the split in the encoding layers and the performance of the model. The more the number of variational layers between the encoding blocks, the better the approximation of the mapping function and accurate the classification in both the training and testing phase. The architecture with ten encoding blocks yields the highest accuracy of around 60\%.\footnote{We acknowledge that the classification accuracy is not competitive for a simple dataset such as MNIST. However, the goal of our work is not the accurate classification of the MNIST dataset but to study the impact of the encoding pattern on the trainability of the model by comparing the relative change in classification accuracy without increasing the number of parameters. Hence, we did not optimize the model for an increased classification accuracy.} We have observed the same pattern on Fashion MNIST. 

To validate the argument that the interval uploading method is not data-dependent and is expected to work on arbitrary classification tasks, we shuffled every pixel value within each image in the MNIST dataset with a fixed permutation chosen randomly. The models trained on this shuffled MNIST dataset also displayed the same pattern as in the other datasets. The test accuracy of these models is shown in~\cref{tab:test_acc_enc}.

\begin{table}
\caption{Interval uploading performance on test sets.}
\label{tab:test_acc_enc}
\centering
\renewcommand{\tabcolsep}{1pt}
\begin{tabular}{p{3.5em}|@{\hspace{1pt}}r|rrrrr} 
Dataset & \multicolumn{1}{c|}{DRU} & \multicolumn{5}{c}{IDU} \\
 &&  \multicolumn{1}{c}{1} & \multicolumn{1}{c}{2} & \multicolumn{1}{c}{4} & \multicolumn{1}{c}{8}  & \multicolumn{1}{c}{10} \\
 \hline 
MNIST         &33.2$\pm$0.01&34.0$\pm$0.01&47.3$\pm$0.03&50.9$\pm$0.01&51.5$\pm$0.00&\textbf{56.7$\pm$0.02}\\
 \textit{shuffled} &32.2$\pm$0.00&47.1$\pm$0.01&52.2$\pm$0.01&53.8$\pm$0.01&56.1$\pm$0.01&\textbf{58.6$\pm$0.01}\\
Fashion&43.5$\pm$0.17&43.8$\pm$0.01&48.3$\pm$0.01&52.5$\pm$0.01&53.6$\pm$0.03&\textbf{56.9$\pm$0.03}
\\
 \end{tabular}
\end{table}

\subsection{IDU in a "Deeper" circuit}

\label{sec:IDU_deeper}
From~\cref{fig:val_enc} and~\cref{tab:test_acc_enc} it becomes clear that the quantum architecture with data re-uploading type encoding exhibits similar or lower performance than the least performing IDU architecture. Intuitively, this is an expected result as the data used for the data re-uploading architecture is a reduced dataset where each image is summed over its columns. This summation results in information loss, hence the loss in performance by the model. However, the architecture with a single encoding layer performs the same summation over the image columns (as only $R_x$ gates are used for encoding) and performs slightly better than the DRU architecture. The poor performance of the model with DRU encoding can be correlated to the low expressive power of the variational layers. A explanation of this hypothesis is given in~\cref{sec:schulz}.

\begin{figure}
    \centering
    \includegraphics[width=\linewidth]{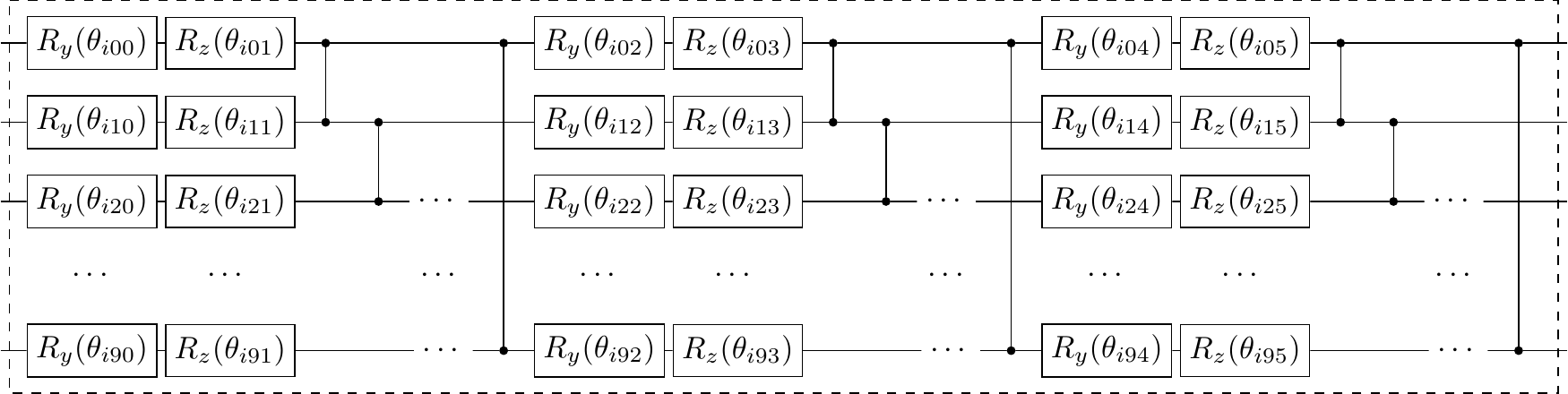}
    \caption{Variational layer with 60 parameters used in deeper models.}
    \label{fig:varialtional_layer_deeper}
\end{figure}

To further validate this hypothesis, we increased the number of parameters in the variational layer from 20 to 60 to increase the trainability of the model, see~\cref{fig:varialtional_layer_deeper}. This in turn increased the performance of the DRU architecture. The accuracy of DRU with a higher number of parameters is better than the architecture with a single encoding layer for the same number of parameters. However, the DRU still demonstrate a significantly low performance compared to all IDU architectures with split greater than two.~\cref{tab:test_acc_enc_deep} shows the results of different architectures with a higher number of parameters.

\begin{table}
\centering
\caption{Performance on a "deeper" architecture.}
\label{tab:test_acc_enc_deep}

\centering
\renewcommand{\tabcolsep}{1pt}
\begin{tabular}{p{3.5em}|@{\hspace{1pt}}r|rrrrr} 
Dataset & \multicolumn{1}{c|}{DRU} & \multicolumn{5}{c}{IDU} \\
 &&  \multicolumn{1}{c}{1} & \multicolumn{1}{c}{2} & \multicolumn{1}{c}{4} & \multicolumn{1}{c}{8}  & \multicolumn{1}{c}{10} \\
 \hline 
MNIST      &42.7$\pm$0.00&41.5$\pm$0.01&54.0$\pm$0.01&57.9$\pm$0.02&62.6$\pm$0.01&\textbf{63.9$\pm$0.01}\\
\end{tabular}
\end{table}

\subsection{IDU for Advanced Encoding schemes}

\begin{figure}[b!]
    \centering
    \includegraphics[width=.95\linewidth]{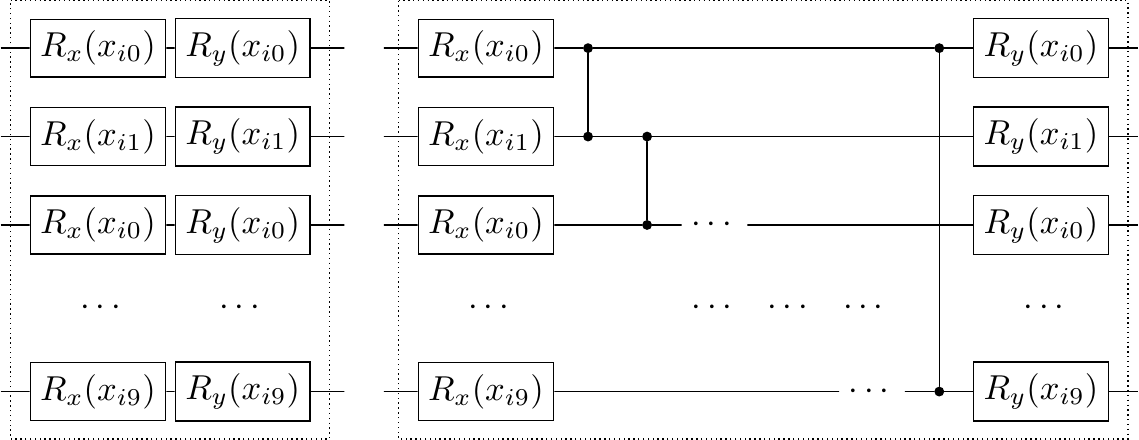}
    \caption{Single encoding block for $R_x$-$R_y$ type encoding (left) and $R_x$-$CR_z$-$R_y$ type encoding (right).}
    \label{fig:advanced_encoding}
\end{figure}

Using single-qubit $R_x$ type encoding gates sequentially results in partial or complete summation of the input image data along the column resulting in some information loss. To further validate the effect of IDU-type encoding pattern without the influence of the summation effect, we tested two other encoding methods: 1) a $R_x$-$R_y$ encoding, where we used a sequence of alternating $R_x$ and $R_y$ rotational gates for each row of the image instead of just $R_x$ gates, see~\cref{fig:advanced_encoding}, and 2) a $R_x$-$CR_z$-$R_y$ encoding, which is similar to $R_x$-$R_y$ encoding but that uses a $CR_z$ gate in between $R_x$ and $R_y$ gates, see~\cref{fig:advanced_encoding}. 

\begin{table}[b]
\caption{Incremental data-uploading performance on advanced encoding methods.}
\label{tab:test_acc_enc_adv}

\centering
\begin{tabular}{p{6.5em}|@{\hspace{1pt}}r|rrrrr} 
Dataset & \multicolumn{1}{c|}{DRU} & \multicolumn{5}{c}{IDU} \\
 &&  \multicolumn{1}{c}{1} & \multicolumn{1}{c}{2} & \multicolumn{1}{c}{4} & \multicolumn{1}{c}{8}  & \multicolumn{1}{c}{10} \\
 \hline 
$R_x$-$R_y$  & 0.23 & 0.29 & 0.34 & 0.37 & 0.43 & \textbf{0.45}\\
$R_x$-$CR_z$-$R_y$  & 0.19 & 0.34 & 0.34 & 0.36 & 0.37 & \textbf{0.41}\\
\end{tabular}
\end{table}

The classification results of MNIST dataset using these two encoding methods are given in~\cref{tab:test_acc_enc_adv}. We see that the effect of incremental data-uploading type encoding pattern is more general and not restricted to single-qubit $R_x$ type encoding. Please note that the encoding methods are simple design choices where the encoding gates do not commute. These encoding methods are not optimized toward the MNIST dataset as the intent behind the experiment was to study the effect of incremental data-uploading on different encoding methods and not the effect of encoding method on MNIST dataset in itself.

%

\section{Theoretical considerations}

\subsection{Quantification of Trainability and Expressibility}\noindent

Two important properties of an \gls{ML} model are its expressibility and trainability. Abbas et al.~\cite{Abbas_2021} generalizes tools for quantitative analysis to the quantum realm. Both concepts are based on the \gls{FIM} \cite{Thomas_2006} associated with the statistical model~\cite{Rissanen_1996} $p_{\theta}(x,y)$ implemented by the \glspl{VQC}. In practice, we use the empirical \gls{FIM} defined as
\begin{equation}
    \Tilde{F}_k(\theta) = \frac{1}{k} \sum_{j=1}^{k} \frac{\partial}{\partial \theta} \ln p_{\theta}(x^{(j)},y^{(j)}) \frac{\partial}{\partial \theta} \ln p_{\theta}(x^{(j)},y^{(j)})^t.
\end{equation}
Here, $(x^{(j)},y^{(j)})_{j=1}^{k}$ are i.i.d. drawn from the joint distribution $p_{\theta}(x,y) = p_{\theta}(y \mid x) p(x)$. For the MNIST dataset one has inputs $x \in \mathbb{R}^{10 \times 10}$ and labels $y \in \left\{ 0, \cdots, 9 \right\}$. However, the following consideration generalize to data of any finite dimensionality.

The \gls{FIM} captures the geometry of the parameter space, which has a crucial influence on the trainability of a model. To assess this, the spectrum of the positive semidefinite matrix, i.e., the distribution of its eigenvalues, is considered. A degenerate spectrum indicates a distorted parameter space, which is disadvantageous for any gradient-based optimization technique. Furthermore, an increasing accumulation of eigenvalues around zero for growing model size (i.e., qubit number) indicates the presence of barren plateaus~\cite{Abbas_2021}.

The effective dimension~\cite{Berezniuk_2020} is a tool to capture the expressibility or capacity of a \gls{ML} model. It is based upon the (empirical) \gls{FIM}, and therefore can be estimated relatively straightforward by sampling. The effective dimension of a statistical model $\mathcal{M}_{\Theta}$ is defined as
\begin{equation}
    ed_n(\mathcal{M}_{\Theta}) := 2 \frac{\ln \left( \frac{1}{V_{\Theta}} \int_{\Theta} \sqrt{\det \left( I_d + c_n \hat{F}(\theta) \right)} d \theta \right)}{\ln \left( c_n \right)},
\end{equation}
where $d = \left| \theta \right|$ is the number of parameters, $V_{\Theta} := \int_{\Theta} d \theta$ is the volume of the parameter space, and $\hat{F}(\theta) \in \mathbb{R}^{d \times d}$ is a normalized version of the (empirical) \gls{FIM}. The parameter $n$ captures the effective resolution of the parameter space (i.e. is related to the data availability). It enters the definition in the normalization factor $c_n = \frac{n}{2 \pi \ln n}$. Under certain conditions the effective dimension provides an upper bound to the generalization error~\cite{Abbas_2021}. In more plain words, the measure quantifies the range of different functions, that a given model can approximate. In order to compare different models, a normalized version of the effective dimension is preferable. A division by $d$ restricts the measure to the range $[0, 1]$, where higher values indicate a more expressible model.

The empirical Fisher information matrix was estimated using 200 random samples $x^k$ from the MNIST data set and 100 random parameter sets $\theta$ of 200 parameters each drawn from a uniform distribution with range $\left[0,\pi \right]$. The eigenvalue spectra of the \glspl{FIM} over 4000 samples for different \gls{IDU} architectures are shown in~\cref{fig:FIM}. For reasons of presentation the histograms cut values larger than one, which anyhow do not change the overall picture. The normalized effective dimension for different IDU architectures for sample sizes ranging from $10^3$ to $10^6$ is shown in~\cref{fig:effDim}. 
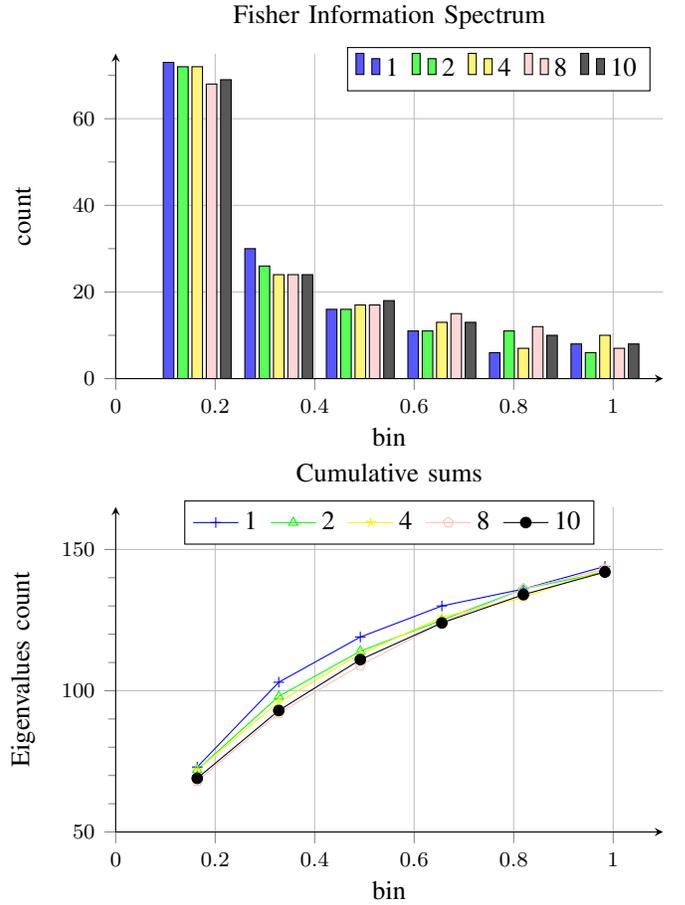
\begin{figure}
\centering
\begin{tikzpicture}
\begin{axis}[name=plot1,title={Fisher Information Spectrum},
width=\linewidth, height=.25\textheight,
grid=major,
tick label style={font=\footnotesize},
minor y tick  num=1,
xmin=0,xmax=1.1,
ymin=0,ymax=75,
ylabel={count},
xlabel={bin},
ybar = .05cm,
bar width = 4pt,
axis x line=bottom, axis y line=left, tick align = outside,
legend columns=-1,
legend style={/tikz/every even column/.append style={column sep=0.1cm},at={(0.7,1)},anchor=south,yshift=-5mm}, 
]
\addplot[fill=blue!66,ybar,no marks] coordinates {
(0.16390732506531325,73)
(0.3278146494044079,30)
(0.49172197374350257,16)
(0.6556292980825973,11)
(0.819536622421692,6)
(0.9834439467607866,8)
};
\addlegendentry{1}
\addplot[fill=green!66,no marks,ybar] coordinates 
{(0.16390732506531325,72)
(0.3278146494044079,26)
(0.49172197374350257,16)
(0.6556292980825973,11)
(0.819536622421692,11)
(0.9834439467607866,6)
};
\addlegendentry{2}
\addplot[fill=yellow!66,no marks,ybar] coordinates 
{(0.16390732506531325,72)
(0.3278146494044079,24)
(0.49172197374350257,17)
(0.6556292980825973,13)
(0.819536622421692,7)
(0.9834439467607866,10)
};
\addlegendentry{4}
\addplot[fill=pink!66,no marks,ybar] coordinates 
{(0.16390732506531325,68)
(0.3278146494044079,24)
(0.49172197374350257,17)
(0.6556292980825973,15)
(0.819536622421692,12)
(0.9834439467607866,7)
};
\addlegendentry{8}
\addplot[fill=black!66,no marks,ybar] coordinates 
{(0.16390732506531325,69)
(0.3278146494044079,24)
(0.49172197374350257,18)
(0.6556292980825973,13)
(0.819536622421692,10)
(0.9834439467607866,8)
};
\addlegendentry{10}

\end{axis}

\begin{axis}[name=plot2,
at=(plot1.below south west), anchor=above north west,
title={Cumulative sums},
width=\linewidth, height=.25\textheight,
ylabel={Eigenvalues count},
xlabel={bin},
xmin=0,xmax=1.1,
ymin=50,
ymax=165,
grid=major,minor y tick  num=4,
tick label style={font=\footnotesize},
axis x line=bottom, axis y line=left, tick align = outside,
legend columns=-1,
legend style={/tikz/every even column/.append style={column sep=0.1cm},at={(0.5,1)},anchor=south,yshift=-5mm}, %
]

\addplot[draw=blue, mark=+] coordinates 
{(0.16390732506531325,73)
(0.3278146494044079,103)
(0.49172197374350257,119)
(0.6556292980825973,130)
(0.819536622421692,136)
(0.9834439467607866,144)
};
\addlegendentry{1}
\addplot[draw=green,mark=triangle] coordinates 
{(0.16390732506531325,72)
(0.3278146494044079,98)
(0.49172197374350257,114)
(0.6556292980825973,125)
(0.819536622421692,136)
(0.9834439467607866,142)
};
\addlegendentry{2}
\addplot[draw=yellow,mark=star] coordinates 
{(0.16390732506531325,72)
(0.3278146494044079,96)
(0.49172197374350257,113)
(0.6556292980825973,126)
(0.819536622421692,133)
(0.9834439467607866,143)
};
\addlegendentry{4}
\addplot[draw=pink,mark=pentagon] coordinates 
{(0.16390732506531325,68)
(0.3278146494044079,92)
(0.49172197374350257,109)
(0.6556292980825973,124)
(0.819536622421692,136)
(0.9834439467607866,143)
};
\addlegendentry{8}
\addplot[draw=black,mark=*] coordinates 
{(0.16390732506531325,69)
(0.3278146494044079,93)
(0.49172197374350257,111)
(0.6556292980825973,124)
(0.819536622421692,134)
(0.9834439467607866,142)
};
\addlegendentry{10}

\end{axis}
\end{tikzpicture}

\caption{Fisher information}
\label{fig:FIM}
\end{figure}
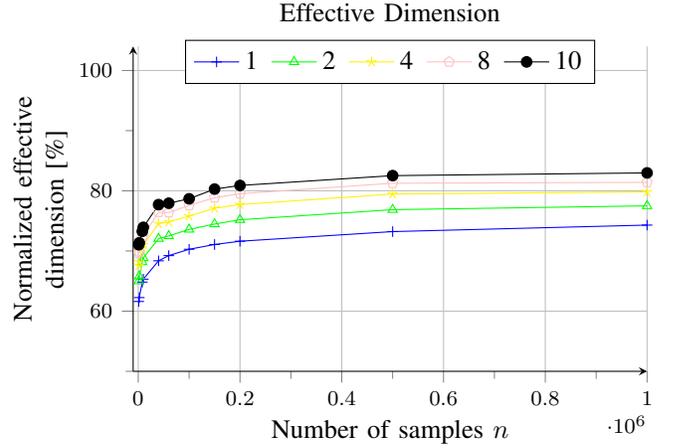
\begin{figure}
\begin{tikzpicture}
\begin{axis}[name=plot1,title={Effective Dimension},
width=.95\linewidth, height=.25\textheight,
ylabel style={align=center}, 
ylabel={Normalized effective\\ dimension [\%]},
xlabel={Number of samples $n$},
xmin=-10000,xmax=1000000,
ymin=50,
ymax=104,
grid=major,
minor y tick  num=1,minor x tick  num=1,
tick label style={font=\footnotesize},
axis x line=bottom, axis y line=left, tick align = outside,
legend columns=-1,
legend style={/tikz/every even column/.append style={column sep=0.1cm},at={(0.5,1)},anchor=south,yshift=-5mm}, %
]

\addplot[draw=blue, mark=+] table[x=n,y=1] {results./compiled/eff_mnist.dat};
\addlegendentry{1}
\addplot[draw=green,mark=triangle] table[ x=n,y=2] {results./compiled/eff_mnist.dat};
\addlegendentry{2}
\addplot[draw=yellow,mark=star] table[ x=n,y=4] {results./compiled/eff_mnist.dat};
\addlegendentry{4}
\addplot[draw=pink,mark=pentagon] table[ x=n,y=8] {results./compiled/eff_mnist.dat};
\addlegendentry{8}
\addplot[draw=black,mark=*] table[ x=n,y=10] {results./compiled/eff_mnist.dat};
\addlegendentry{10}

\end{axis}
\end{tikzpicture}

\caption{Effective dimension for different IDU architectures.}
\label{fig:effDim}
\end{figure}
Results presented in~\cref{fig:FIM} depict that the eigenvalue spectrum becomes more uniform for IDU architectures with a higher number of splits. This normalizing effect becomes more obvious when considering the cumulative sum plot shown in~\cref{fig:FIM}, bottom.

Although the peculiarity of the normalization effect is quite small in the considered instances, it indicates an improvement in trainability when employing the proposed approach. As the spectrum is more uniform with fewer eigenvalues close to zero, the parameter space is less distorted, which is beneficial for optimization methods. The difference in terms of the effective dimension is more distinct, i.e. it clearly increases when using more IDU layers. This indicates an increase in model expressibility, while the number of parameters stays the same. In all instances the normalized effective dimension grows with larger resolution of the parameter space, which is a reasonable behaviour for machine learning models.

\subsection{Frequency spectrum}\noindent
\label{sec:schulz}

To gain more insight into the performance differences observed in the previous sections, in the following we investigate the function class represented by the different architectures. Slightly generalizing the setting, we consider $x \in \mathbb{R}^{N \times M}$ in the following and denote the $j$th rows of the matrix by the column vector $\bm{x}_j$, that is ($\bm{x}_j)_k=x_{jk}$ for $k=0,...,M-1$ and $j=0,...,N-1$.
The vectors $\bm{x}_j$ therefore correspond to the data fed in the $j$th encoding layer in~\cref{fig:overview}. 
It was shown by Schuld et al.\ in Ref.~\cite{Schuld_2021} that the functions $f_\theta$ represented by VQCs are Fourier sums when each encoding layer is given by single-qubit rotations about a given axis for each qubit. In particular, the variational layers determine the amplitudes and the frequency spectrum is fixed by the data-encoding layers. Following Ref.~\cite{Schuld_2021}, we find
\begin{equation}
\label{eq:Fouriersum}
   f_\theta (x) =\sum_{\bm{\omega}_0,...,\bm{\omega}_{N-1} \in \Omega} c_{\omega}(\theta) \mathrm{exp}\left\{i \sum_{j=0}^{N-1} \bm{\omega}_j  \bm{x}_j \right\}\,,
\end{equation}
where $\Omega=\{-1,0,1\}^M$ is the frequency spectrum and $\omega$ the matrix containing $\bm{\omega}_j$ as $j$th row. Since $f_\theta$ is real valued, we find $c_{-\omega}=c^*_{\omega}$. More intuitively, the functions represent $NM$ dimensional Fourier sums with frequencies $\pm 1$ and $0$. Note that the coefficients $c_\omega (\theta)$ are only independent and can be chosen freely if the variational layers are universal, i.e.~can represent any unitary matrix. In practice, the expressivity of the circuit might be severely limited by the number of variational parameters, indeed a general $n$-qubit unitary requires exponentially many parameters in the numbers of qubits. Nevertheless, equation (\cref{eq:Fouriersum}) qualitatively explains the behaviour shown in~\cref{tab:test_acc_enc} where an increase of performance with increasing number of interleaved variational layers is observed. Since the data encoding is based on $R_x$ rotations only, the setup in the left subfigure of~\cref{fig:splits} is equivalent to summing the vectors and feeding the result into the circuit by only one encoding layer. As a result, the input dimension in equation (\cref{eq:Fouriersum}) decreases from $NM$ to $M$ so that the model loses access to much of the information present in the data $x$, explaining the poor performance in the left column of~\cref{tab:test_acc_enc}. As the number of variational layers increase, $f_\theta$ gains access to more information as only some of the rows in the data are summed, finally reaching optimal performance for the fully interleaved setup shown in the right subfigure of \cref{fig:splits}. It is worthwhile noting that decreasing the number of interleaved layers while keeping the number of variational layers constant, increases the expressibility of the final variational layers but it
seems conceivable that this increase cannot compensate for the information loss by partially or fully summing the rows in the data $x$. The same argument applies to the interpretation of the 
DRU column in \cref{tab:test_acc_enc} and \cref{tab:test_acc_enc_deep}. Here, in the data re-uploading setting the rows of the image are first summed and then repeatedly encoded into the circuit with variational layers in between. While the frequency spectrum of the Fourier sum now contains all integers between $-N$ and $N$ \cite{Schuld_2021}, as can be seen from equation (\cref{eq:Fouriersum}) by replacing $\bm{x}_j$ by the sum over the rows for all $j$, again the model seems unable to compensate for the information which is lost in summing the rows of the image.
In case of more general encoding schemes such as alternating $R_x$-$R_y$ gates for subsequent encoding layers, intuitively, this effect is less dramatic due to the non-commutativity of the encoding gates. However, \cref{tab:test_acc_enc_adv}  indicates that also in case of non-commuting encoding gates the information in the data is much better accessible by the model in which the more variational layers are interleaved with endoding layers. A more rigorous discussion of this situation will be at the focus of further work.

\section{Conclusion}

This paper proposes an encoding pattern called incremental data-uploading for high dimensional data. It acts as a guideline for positioning the encoding layers in a variational quantum circuit. Here, the encoding and variational layers alternate one after the other so that the data is fed incrementally into the circuit and becomes more accessible to the model.
IDU with maximum variational layers in between them showed a performance boost of 15 - 25 percentage points in image classification tasks. The effective dimension and Fisher information results also support our claim that the IDU pattern increases the trainability and expressivity of the QML model without increasing the number of its parameters. In addition, expressing the quantum model as a partial Fourier sum, we were able to connect its performance to the range of accessible frequencies.

Our experiments also showed that an encoding pattern like data re-uploading exhibits low accuracy when dealing with high dimensional data and fewer parameters in a QML model even though it increases the overall QC depth and number of quantum gates. Hence, we conclude that the presented data encoding pattern shows an improvement in the performance of a QML model with high dimensional data, shallow circuit depth and a given encoding method. However, finding an optimal encoding method in the IDU framework for a given dataset is left for future work.

\bibliographystyle{IEEEtran}
\bibliography{references}

\end{document}